\documentclass[preprint,showpacs,preprintnumbers,amsmath,amssymb]{revtex4}

\usepackage{graphicx}
\usepackage{dcolumn}
\usepackage{bm}

\begin{document}

\preprint{}

\title{Volumes and Hyperareas of the Spaces of 
Separable and Nonseparable Qubit-{\it Qutrit} Systems: Initial Numerical Analyses}

\author{Paul B. Slater}%
\email{slater@kitp.ucsb.edu}
\affiliation{%
ISBER, University of California, Santa Barbara, CA 93106\\
}%
\date{\today}

\begin{abstract}
Paralleling our recent computationally-intensive work for the case $N=4$
(quant-ph/0308037),
we undertake the task for $N=6$ of computing to
high numerical accuracy, the formulas
of Sommers and \.Zyczkowski (quant-ph/0304041) for the $(N^2-1)$-dimensional volume and $(N^2-2)$-dimensional 
hyperarea of the (separable  and nonseparable) 
$N \times N$ density
matrices, based on the Bures (minimal monotone) metric. At the same time, 
we estimate the {\it unknown} volumes and hyperareas based on 
 a number of other monotone metrics of interest. Additionally, we 
estimate --- but 
with perhaps unavoidably diminished
accuracy --- all these volume and hyperarea quantities, when restricted
to the ``small'' subset of 
$6 \times 6$ density matrices that are separable (classically correlated) in nature. The ratios of separable to separable plus nonseparable volumes, then,
yield corresponding estimates of the ``probabilities of separability''.
We are particularly interested in 
exploring 
the possibility that a number of the various 35-dimensional volumes and 34-dimensional hyperareas, possess {\it exact} 
values --- which we had, in fact, conjectured to be the case for the qubit-qubit systems ($N=4$), with the ``silver mean'', $\sqrt{2}-1$, appearing to play
a fundamental role as regards the separable states.
\end{abstract}

\maketitle

\section{Introduction}
In a recent highly comprehensive 
analysis \cite{hans1} (cf. \cite{hansrecent}), Sommers and \.Zyczkowski obtained
``a fairly general expression for the Bures volume of the submanifold of the states of rank $N-n$ of the set of complex ($\beta=2$) or real ($\beta=1$)
$N \times N$ density matrices
\begin{equation} \label{HZ1}
S^{(\beta)}_{N,n}= 2^{-d_{n}} \frac{\pi^{(d_{n}+1)/2}}{\Gamma((d_n+1)/2)} \Pi^{N-n}_{j=1} \frac{\Gamma(j \beta/2) \Gamma[1+(2 n+j-1) \beta/2]}{\Gamma[(n+j) \beta/2] \Gamma[1+(n+j-1) \beta/2]},
\end{equation}
where $d_{n}= (N-n) [1+(N+n-1) \beta/2] -1$ represents the dimensionality of the
manifold \ldots for $n=0$ the last factor simply equals unity and (\ref{HZ1})
gives the Bures volume of the entire space of density matrices, equal to that
of a $d_{0}$-dimensional hyper-hemisphere with radius 1/2. In the case
$n=1$ we obtain the volume of the surface of this set, while for $n=N-1$ we get the volume of the set of pure states \ldots which for $\beta=1(2)$ gives correctly the volume of the real (complex) projective space of dimensions
$N-1$'' \cite{hans1}.

The Bures metric on various spaces of density matrices has been widely
studied \cite{hubner1,hubner2,ditt1,ditt2}. In a broader context, it serves
as the {\it minimal} monotone metric \cite{petz1}.

Let us apply (\ref{HZ1}) to the cases of specific interest in this study,
$N=6,n=0,\beta=2$ and $N=6,n=1,\beta=2$ --- that is, the Bures 35-dimensional volume and 
34-dimensional hyperarea of the complex $6 \times 6$ density matrices.
We then have that 
\begin{equation} \label{m1}
S^{(2)}_{6,0}= \frac{{\pi }^{18}}{12221326970165372387328000} \approx 
7.27075 \cdot {10}^{-17}
\end{equation}
and
\begin{equation} \label{m2}
S^{(2)}_{6,1}= \frac{{\pi }^{17}}{138339065763438059520000} \approx 2.04457 \cdot {10}^{-15}.
\end{equation}
Here,  we are able (somewhat paralleling our recent work for $N=4$ 
\cite{slatersilver}, but in a 
rather
more systematic manner than there) through numerical (quasi-Monte Carlo/quasi-random) methods to reproduce 
both of these values (\ref{m1}), (\ref{m2}), to a considerable accuracy.
At the same time, we compute numerical values --- it would seem reasonable
to assume, 
with roughly the same level of accuracy --- of these two quantities, but for the replacement of the Bures metric by five other {\it monotone} metrics of
interest. These are the Kubo-Mori \cite{hasegawa,petz3,michor,streater},
(arithmetic) average \cite{slatersilver}, Wigner-Yanase \cite{gi,wy,luo,luo2},
 Grosse-Krattenthaler-Slater  (GKS) \cite{KS,gillmassar} 
 and (geometric) average monotone 
metrics --- the two ``averages'' being formed from the minimal and {\it maximal} monotone metrics, following the suggested procedure in
\cite[eq. (20)]{petz2}. No proven 
formulas, such as (\ref{HZ1}), are presently available for these various quantities, 
although our research here and in \cite{slatersilver} strongly suggests that
the Kubo-Mori volume of the $N \times N$ density matrices is equal 
to simply $2^{N(N-1)/2)} S^{(2)}_{N,0}$, which for our case of $N=6$ would be  $32768 S^{(2)}_{6,0}$ (cf. Tables I and II).
(In light of the considerable attention recently devoted to the (Riemannian, but non-monotone) Hilbert-Schmidt metric \cite{hansrecent,hilb1,hilb2} --- it would 
certainly seem appropriate to include it as
well in any further analyses along the lines here and in \cite{slatersilver}.)

Further, we compute for all these six metrics the 35-dimensional volumes and
34-dimensional hyperareas restricted to the {\it separable} $2 \times 3$ and
$3 \times 2$ systems. (Then, we can, obviously, by taking ratios obtain ``probabilities of separability'' --- a topic which was first investigated in \cite{ZHSL}, 
and studied further, using the Bures metric, in \cite{zycz2,slaterA,slaterC}).
For this purpose, we employ the convenient Peres-Horodecki necessary {\it and} 
sufficient positive 
partial
transposition criterion for separability \cite{asher,michal} --- asserting that a $4 \times 4$ or $6 \times 6$ density matrix is separable if and only if all the eigenvalues of its
partial transpose are positive. (But in the $6 \times 6$ case, we have the qualitative difference that partial transposes can be determined in two inequivalent ways, either by transposing in place, in the natural
manner, 
the nine $2 \times 2$ submatrices or the four 
$3 \times 3$ submatrices. We will throughout this study --- as in \cite{qubitqutrit} --- analyze results using {\it both} forms of partial transpose.
It is our anticipation --- although without a formal demonstration --- that in the limit of large sample size, the two sets of results will converge
to true {\it common} values.)

Our main analysis takes the form of a quasi-Monte Carlo (Tezuka-Faure 
\cite{tezuka})  
numerical integration over the 35-dimensional hypercube ($[0,1]^{35}$) and a 
34-dimensional subhypercube of it. In doing so, we implement 
 a parameterization of the $6 \times 6$ density matrices in terms of thirty Euler angles (parameterizing $6 \times 6$ unitary matrices) and {\it five} hyperspherical angles
(parameterizing the {\it six} eigenvalues --- constrained to sum to 1 
\cite{sudarshan,toddecg}). We hold a single one of the five hyperspherical angles fixed in the 34-dimensional analysis.  (The parameters are linearly transformed so 
that they all lie in the interval [0,1].)

We have previously pursued a similar numerical analysis in investigating the separable and nonseparable volumes and hyperareas of the $4 \times 4$ density matrices \cite{slatersilver}. 
Highly accurate results (as gauged in terms of {\it known} 
Bures quantities \cite{hans1}) --- based on two {\it billion} points of a Tezuka-Faure (``low discrepancy'') 
sequence lying in the 15-dimensional hypercube --- led us to advance several strikingly simple conjectures.
For example, it was indicated  that the Kubo-Mori volume of separable and nonseparable states was exactly $64 =2^6$ times the Bures volume. (The exponent 
6 is  expressible --- in terms of our general conjecture, 
mentioned above,  
relating the Bures and Kubo-Mori volumes --- as $N(N-1)/2$, $N=4$.)
Most prominently, though, it appeared that the statistical distinguishability 
(SD) volume 
was simply expressible as $\frac{\sigma_{Ag}}{3}$, where the ``silver mean''
\cite{christos,spinadel,gumbs,kappraff},
$\sigma_{Ag} = \sqrt{2}-1 \approx 0.414214$, and $10 \sigma_{Ag}$ in terms of 
(four times) the Kubo-Mori
metric. (The SD metric is identically four times the Bures metric
\cite{caves}. Consequently, the SD 15-dimensional volume of the $4 \times 4$ complex density matrices
is $2^{15}$ times that of the Bures volume --- given by (\ref{HZ1}) for $N=4,n=0,\beta=2$ --- thus equalling the volume
of a 15-dimensional hyper-hemisphere with radius 1, rather than $\frac{1}{2}$
 as in the Bures case itself.)
Unfortunately, there appears to be little in the way of ``clues'' in the literature, as to how one might {\it formally} prove or disprove these conjectures --- ``brute
force'' {\it symbolic} integration appearing to be well beyond present technical/conceptual capabilities --- although the author ``suspects'' that at least in the Bures/{\it minimal} 
monotone case, a proof might conceiveably be based on the concept of ``minimal volume'' \cite{bayard,bowditch,bambah}. (Certainly, Sommers and \.Zyczkowski \cite{hans1} did not directly
employ symbolic integration methodologies in deriving the Bures volume, hyperarea...for $N$-level [separable {\it and}  nonseparable] systems, but rather, principally, 
used concepts of random matrix theory.)

The monotone metrics (of which we study five, in addition to the
Bures) can all be expressed
in the form
\begin{equation}
g_{\rho}(X',X) 
 = \frac{1}{4} \Sigma_{\alpha,\beta} |\langle \alpha |X| \beta \rangle |^2 c_{monotone}(\lambda_{\alpha},\lambda_{\beta})
\end{equation}
(cf. \cite{hubner1,hubner2}).
Here $X,X'$ lie in the tangent space of all Hermitian $N \times N$ density
matrices $\rho$ and $|\alpha \rangle, \alpha =1, 2 \ldots$ are eigenvectors
of $\rho$ with eigenvalues $\lambda_{\alpha}$.
Now, $c_{monotone}(\lambda_{\alpha},\lambda_{\beta})$ represents the specific {\it Morozova-Chentsov} function for the monotone metric in question \cite{petz2}. This function takes the form
for the Bures metric,
\begin{equation} \label{Bures}
c_{Bures}(\lambda_{\alpha},\lambda_{\beta}) = \frac{2}{\lambda_{\alpha} +\lambda_{\beta}},
\end{equation}
for the Kubo-Mori metric (which, up up to a scale factor, is 
the unique monotone
Riemannian metric with respect to which the exponential and mixture
connections are dual \cite{streater}),
\begin{equation} \label{KM}
c_{KM}(\lambda_{\alpha},\lambda_{\beta}) =\frac{\log{\lambda_{\alpha}}-\log{\lambda_{\beta}}}{\lambda_{\alpha}-\lambda_{\beta}},
\end{equation}
for the (arithmetic) average metric (first discussed in \cite{slatersilver}),
\begin{equation}
c_{arith}(\lambda_{\alpha},\lambda_{\beta}) = \frac{4 (\lambda_{\alpha}+\lambda_{\beta})}{\lambda_{\alpha}^2 + 6 \lambda_{\alpha} \lambda_{\beta} +\lambda_{\beta}^2},
\end{equation}
for the Wigner-Yanase metric (which corresponds
to a space of {\it constant curvature} \cite{gi}),
\begin{equation}
c_{WY}(\lambda_{\alpha},\lambda_{\beta}) =\frac{4}{(\sqrt{\lambda_{\alpha}} +\sqrt{\lambda_{\beta}})^2},
\end{equation}
for the GKS/quasi-Bures metric (which yields the asymptotic redundancy for
universal quantum data compression \cite{KS}),
\begin{equation}
c_{GKS}(\lambda_{\alpha},\lambda_{\beta})= \frac{{\frac{\lambda_{\alpha}}{\lambda_{\beta}}}^{\lambda_{\alpha}/(\lambda_{\beta}-\lambda_{\alpha})}}{\lambda_{\beta}} e
\end{equation}
and for the (geometric) average metric (apparently previously unanalyzed),
\begin{equation} \label{geom}
c_{geom}(\lambda_{\alpha},\lambda_{\beta}) =\frac{1}{2 \sqrt{ \lambda_{\alpha} \lambda_{\beta}}}.
\end{equation}
\section{Analyses}
Based on the first 600 million points of a Tezuka-Faure sequence, to which we are continuing to copiously add, we 
obtained the results reported in Tables~\ref{tab:table1}-\ref{tab:table7}. (We followed the Bures 
formulas in \cite[secs. III.C, III.D]{hans1}, substituting the Morozova-Chentsov functions
given above (\ref{KM})-(\ref{geom}), 
in the appropriate manner, to obtain their counterparts for the various non-Bures monotone metrics.)

In Table~\ref{tab:table1}, we scale the estimates of the volumes and hyperareas by the {\it known} values (\ref{m1}), (\ref{m2}) of
$S^{(2)}_{6,0}$ and $S^{(2)}_{6,1}$, while in Table~\ref{tab:table2} we scale these estimates by the {\it estimated} values ($7.21259 \cdot 10^{-17}$  and $2.04607 \cdot 10^{-15}$) of these two quantities.
(We use {\it both} approaches because we are uncertain as to which may be more revealing as to possible exact ratios --- the possibility of which is
suggested by our work in \cite{slatersilver}. It is interesting to observe that the convergence to the true values of $S^{(2)}_{6,0}$ and $S^{(2)}_{6,1}$ appears
to be more pronounced in the 34-dimensional case than in the 35-dimensional one, although the Tezuka-Faure sequence we employ is specifically designed
as a {\it 35}-dimensional one --- of which we take an essentially arbitrary
34-dimensional {\it projection} (cf. \cite[sec. 7]{morokoff}).)

\begin{table}
\caption{\label{tab:table1}Scaled estimates based on the Tezuka-Faure sequence of 600 million points of 
 the 35-dimensional volumes and 34-dimensional
hyperareas of the $6 \times 6$ density matrices, using several monotone metrics. The scaling factors are the {\it known}
values of the volume and 
hyperarea for the Bures metric, given by (\ref{HZ1}), and more specifically
for the cases $N=6$,  $n=0,1$, $\beta=2$ by
(\ref{m1}) and (\ref{m2}).}
\begin{ruledtabular}
\begin{tabular}{rrr}
metric & volume/$S_{6,0}^{(2)}$ &  hyperarea/$S_{6,1}^{(2)}$ \\
\hline
Bures & 0.992001  & 1.00073\\
KM & 31046.8  & 45.5328 \\
arith & 614.789  & 31.3225 \\
WY & 130.323  & 9.78041 \\
GKS & 12.2984  & 3.56433\\
geom & $6.53456 \cdot 10^{38}$ & $2.1581 \cdot 10^{9}$  \\
\end{tabular}
\end{ruledtabular}
\end{table}
\begin{table}
\caption{\label{tab:table2}Scaled estimates based on the Tezuka-Faure sequence of 600 million points of the 35-dimensional volumes and 34-dimensional
hyperareas of the $6 \times 6$ density matrices, using several monotone metrics. The scaling factors ($\tilde{S}$) are the {\it estimated}
values ($7.21259 \cdot 10^{-17}$  and \newline 
$2.04606 \cdot 10^{-15}$ ) of the volume and hyperarea for the Bures metric.}
\begin{ruledtabular}
\begin{tabular}{rrr}
metric &  volume/$\tilde{S}^{(2)}_{6,0}$  &  
hyperarea/$\tilde{S}^{(2)}_{6,0}$ \\
\hline
KM & 31297.2 & 45.4995\\
arith & 619.747 & 31.2995\\
WY & 131.374 & 9.77326\\
GKS & 12.3976 & 3.56172 \\
geom & $6.58726 \cdot 10^{38}$ & $2.15652 \cdot 10^{9}$ \\
\end{tabular}
\end{ruledtabular}
\end{table}
In Tables~\ref{tab:table3} and Tables~\ref{tab:table4}, we report our estimates (scaled by the values obtained for the Bures metric) 
of the volumes and hyperareas of the $6 \times 6$ separable complex density matrices.
Let us note, however, that 
to compute the hyperarea of the complete boundary of the separable states, one must also include those $6 \times 6$ density matrices of full rank,
the partial transposes of which have a zero eigenvalue, and all other eigenvalues  nonnegative 
 \cite{shidu}. (We do not compute this contribution here, as it would slow considerably 
the overall process in which we are engaged, since high-degree polynomials
would need to be 
solved at each step.) In \cite{slatersilver}, we had been led to conjecture that
that part of the 14-dimensional boundary of separable $4 \times 4$ density matrices consisting
generically of rank-{\it four} density matrices had 
SD hyperarea $\frac{55 \sigma_{Ag}}{39}$
and that part composed of rank-{\it three} density matrices, $\frac{43 \sigma_{Ag}}{39}$, for a 
total 14-dimensional boundary 
SD hyperarea of $\frac{98 \sigma_{Ag}}{39}$. We, then, sought to apply the
``Levy-Gromov isoperimetric inequality'' to the relation between the 
known and estimated SD
volumes and hyperareas of the separable and separable plus nonseparable
states \cite[sec. VII.C]{slatersilver}.

In Table~\ref{tab:table3} we compute the partial transposes of the $6 \times 6$ density matrices by transposing in place the four  $3 \times 3$ submatrices, while
in Table~\ref{tab:table4} we transpose in place the nine 
 $2 \times 2$ submatrices.
\begin{table}
\caption{\label{tab:table3}Scaled estimates based on the Tezuka-Faure sequence of 600 million points of the 35-dimensional volumes and 34-dimensional
hyperareas of the {\it separable} $6 \times 6$ density matrices, using several monotone metrics. The scaling factors are the {\it estimated}
values ($1.58803 \cdot 10^{-19}$ and $1.86401 \cdot 10^{-18}$) --- the true values being unknown --- of the volume  and hyperarea  for the Bures metric. To implement the Peres-Horodecki positive partial transposition criterion,
 we compute
the partial transposes 
of the four $3 \times 3$ submatrices (blocks) of the density matrix.}
\begin{ruledtabular}
\begin{tabular}{rrr}
metric & Bures-scaled volume  &  Bures-scaled hyperarea  \\
\hline
KM & 2491.77 & 9.83988\\
arith & 226.332 & 11.3251\\
WY & 49.7186 & 4.30152 \\
GKS & 8.05581 & 2.35047\\
geom & $3.66418 \cdot 10^{23}$ & 52086.9 \\
\end{tabular}
\end{ruledtabular}
\end{table}
\begin{table}
\caption{\label{tab:table4}Scaled estimates based on the Tezuka-Faure sequence of 600 million points of the 35-dimensional volumes and 34-dimensional
hyperareas of the {\it separable} $6 \times 6$ density matrices, using several monotone metrics. The scaling factors are the {\it estimated}
values ($1.03484 \cdot 10^{-19}$  and $1.46223 \cdot 10^{-18}$) --- the true values being unknown --- of the volume and hyperarea 
for the Bures metric. To implement the Peres-Horodecki positive partial transposition criterion, we compute
the partial transposes of the nine  $2 \times 2$ submatrices (blocks)  of the density 
matrix.}
\begin{ruledtabular}
\begin{tabular}{rrr}
metric & Bures-scaled volume &  Bures-scaled hyperarea  \\
\hline
KM & 2563.64 & 8.41556 \\
arith & 198.163 & 10.3172 \\
WY & 46.5238 & 3.98471 \\
GKS & 7.59133 & 2.25601 \\
geom & $7.61764 \cdot 10^{23}$ & 12275.1 \\
\end{tabular}
\end{ruledtabular}
\end{table} 
In Table~\ref{tab:table5}, we only require the density matrix 
in question to pass {\it either} of the two tests, while 
in Table~\ref{tab:table6}, we require it 
to pass {\it both} tests for separability.
(Of the 600 million points of the Tezuka-Faure 35-dimensional sequence so 
far generated, 
approximately 2.91 percent yielded density matrices passing the test for Table I, 2.84 percent  for
Table II, 4 percent  for Table III and 1.74 percent  for Table IV.)
\begin{table}
\caption{\label{tab:table5}Scaled estimates based on the Tezuka-Faure sequence of 600 million points of the 35-dimensional volumes and 34-dimensional
hyperareas of the {\it separable} $6 \times 6$ density matrices, using several monotone metrics. The scaling factors are the {\it estimated}
values ($2.59687 \cdot 10^{-19}$  and $3.30035 \cdot 10^{-18}$)--- the true values being unknown --- for the Bures metric. 
A density matrix is included here if it passes {\it either} form of the positive partial transposition test.}
\begin{ruledtabular}
\begin{tabular}{rrr}
metric & Bures-scaled volume  & Bures-scaled  hyperarea  \\
\hline
KM & 2534.46 & 9.25073\\
arith & 216.379 & 10.9214 \\
WY & 48.6732 & 4.17378 \\
GKS & 7.89451 & 2.31228  \\
geom & $4.63187 \cdot 10^{23}$ & 34855.9 \\
\end{tabular}
\end{ruledtabular}
\end{table}
\begin{table}
\caption{\label{tab:table6}Scaled estimates based on the Tezuka-Faure sequence of 600 million points of the 35-dimensional volumes and 34-dimensional
hyperareas of the {\it separable} $6 \times 6$ density matrices, using several monotone metrics. The scaling factors are the {\it estimated}
values ($2.59945 \cdot 10^{-21}$  and $3.59687 \cdot 10^{-31}$) --- the true values being unknown --- for the Bures metric. 
A density matrix is included here {\it only} 
if it passes 
{\it both} forms of the positive partial transposition test.}
\begin{ruledtabular}
\begin{tabular}{rrr}
metric & Bures-scaled volume  &  Bures-scaled hyperarea  \\
\hline
KM & 1087.9 & 4.50082\\
arith & 98.6297 & 5.85995 \\
WY & 26.9706 & 2.69208 \\
GKS & 5.67912  & 1.80686 \\
geom & $6.43785 \cdot 10^{24}$ & 117.375 \\
\end{tabular}
\end{ruledtabular}
\end{table}

In Table VII, we ``pool'' (average) the results for the separable volumes
and hyperareas reported in Tables III and IV, based on the two 
distinct forms of
partial transposition, to obtain possibly superior estimates of these
quantities, which presumably are actually one and the same {\it independent}
of the particular form of partial transposition.
\begin{table}
\caption{\label{tab:table7}Scaled estimates 
obtained by pooling the results from Tables III and
IV --- based on the two forms of partial transposition --- for 
the separable volumes and hyperareas. The Bures scaling factors
(pooled volume and hyperarea) are $1.31143 \cdot 10^{-19}$ 
and $1.66312 \cdot 10^{-18}$}
\begin{ruledtabular}
\begin{tabular}{rrr}
metric & Bures-scaled volume & Bures-scaled hyperarea \\
\hline
Bures & 2520.12 & 9.21374 \\
arith & 215.212 & 10.882 \\
WY & 48.4581 & 4.16225 \\
GKS & 7.87255 & 2.30894 \\
geom & $5.224 \cdot 10^{23}$ & 34585.5 \\
\end{tabular}
\end{ruledtabular}
\end{table}
\section{Discussion}
Of course, by taking the {\it ratios} of estimates of the volumes/hyperareas
 of separable states to the estimates of the volumes/hyperareas of
separable plus nonseparable states, one would, in turn,  obtain estimates of
the probabilities of separability \cite{ZHSL} 
for the various monotone metrics studied. (Obviously, scaling the estimated volumes
and hyperareas by the corresponding estimates for the  
Bures metric, as we have done in the
tables above for numerical convenience and possible insightfulness,
would be inappropriate in such a process.) The Bures metric gives the
{\it largest} probability of separability.

In \cite{qubitqutrit}, we attempted a somewhat similar quasi-Monte Carlo 
qubit-qutrit 
analysis (but restricted to the Bures metric) to that reported above, but based on many fewer points (70 million {\it vs.} the 600 million so far 
used here) of a 
(Halton) sequence.
At this stage, having made use of considerably increased computer power (and streamlined MATHEMATICA programming --- in particular 
employing the Compile command, which enables the program to 
proceed under the condition
that certain variables will enter a calculation only as 
machine numbers, and
not as lists, algebraic objects or any other kind of expression),
we must regard this earlier study as superseded by the one here.
(Our pooled estimate of the Bures volume of the separable qubit-qutrit
systems here  [Table VII] 
is $1.31143 \cdot 10^{-19}$, while in \cite{qubitqutrit}, following our earlier work for $N=4$ \cite{firstqubitqubit}, 
we formulated a conjecture (\cite[eq. (5)]{qubitqutrit}) --- in which we can now have but 
very little
confidence --- that would give [converting from the SD metric to the 
Bures] a value of
 $2.19053 \cdot 10^{-9} \cdot 2^{-35} \approx  6.37528 \cdot 10^{-20}$.)
We also anticipate revisiting the $N=4$ (qubit-qubit) case \cite{slatersilver}
with our newly accelerated programming methods.

We continue to add to the 600 million points of the Tezuka-Faure sequence employed above, and hope to report considerably more accurate results in 
the future (based on which, hopefully, 
 we can advance plausible hypotheses as to
 the true underlying 
values of the 35-dimensional volumes and 
34-dimensional hyperareas). In fact, at the time of submission of this paper, we have already generated an additional 110 million points, using several {\it
independent} processors. Since the additional points are not numbered 600,000,0001 to 710,000,000 in the  
sequence --- with the gaps remaining to be filled in --- it seems inappropriate to report the results  fully now. Let us only indicate that if we were to do so, 
the first line of Table I (that is, 0.992001=1/1.00806  and 1.00073=1/0.999269) would be replaced
by 0.9985  and 0.999644, so certainly the prognosis for considerably greater accuracy, as we extend the length of the sequence, 
is good.

It would be interesting to conduct analogous investigations to those reported here and in \cite{slatersilver} for the case $N=4$, using
quasi-random sequences {\it other}
 than Tezuka-Faure ones \cite{tezuka}, particularly those for which it is possible to do {\it statistical} testing on the results
(such as constructing confidence intervals) \cite{hong}. (It is, of course, 
possible to conduct statistical testing using simple Monte Carlo methods, but
their convergence is much slower than that of the quasi-Monte Carlo
procedures. Since we are dealing with quite high-dimensional spaces, good
convergence has been our dominant consideration in the selection of
numerical integration methodologies to employ.)

\begin{acknowledgments}
I wish to express gratitude to the Kavli Institute for Theoretical
Physics for computational support in this research and to Giray \"Okten
for supplying the MATHEMATICA code for the Tezuka-Faure quasi-Monte Carlo
procedure.

\end{acknowledgments}

\bibliography{QQ}

\end{document}